\documentclass[nofootinbib,twocolumn,showpacs,superscriptaddress,twoside,prx]{revtex4-1}
\usepackage{amsmath}
\usepackage{amssymb}
\usepackage{amsfonts}
\usepackage{enumerate}
\usepackage{mathrsfs}
\usepackage{graphicx}
\usepackage{epstopdf}
\usepackage{enumerate}
\usepackage{changepage}
\usepackage{diagbox}
 
\usepackage{booktabs}

\newtheorem{theorem}{Theorem}
\newtheorem{definition}{Definition}

\newtheorem{lemma}{Lemma}
\newtheorem{proposition}{Proposition}
\newtheorem{example}{Example}

\usepackage[
colorlinks,
linkcolor = blue,
citecolor = blue,
urlcolor = blue]{hyperref}
\def \qed {\hfill \vrule height7pt width 7pt depth 0pt}

\begin{document}

\title{ Absolutely entangled set of pure states}

\author{Mao-Sheng Li}
\email{li.maosheng.math@gmail.com}
\affiliation{Department of Physics, Southern University of Science and Technology, Shenzhen, 518055, China}
	\affiliation{ Department of Physics, University of Science and Technology of China, Hefei, 230026, China}

\author{Man-Hong Yung}
\email{yung@sustc.edu.cn}
\affiliation{Department of Physics, Southern University of Science and Technology, Shenzhen, 518055, China}
\affiliation{Institute for Quantum Science and Engineering, and Department of Physics,
Southern University of Science and Technology, Shenzhen, 518055, China}

\begin{abstract}
  Quite recently,  Cai et al. [\href{http://arxiv.org/abs/arXiv:2006.07165v1}{arXiv:2006.07165v1}] proposed a new concept ``absolutely
entangled set" for bipartite quantum systems: for any possible choice of global basis, at least one   state
of the set is entangled.  There they  presented a minimum example  with a set of four states in two qubit systems and they proposed a quantitative measure for the absolute set entanglement. In this work, we derive two necessity conditions for a set of states to be   an absolutely
entangled set. In addition,  we give a series constructions of   absolutely entangled bases on $\mathbb{C}^{d_1}\otimes \mathbb{C}^{d_2}$ for any nonprime dimension $d=d_1\times d_2$.   Moreover, based on the structure of the orthogonal product basis in $\mathbb{C}^2\otimes \mathbb{C}^n$, we obtain another construction of absolutely entangled set with $2n+1$ elements in $\mathbb{C}^2\otimes \mathbb{C}^n$.
\end{abstract}

\maketitle

\section{Introduction}

The phenomenon of entanglement \cite{Horodecki09}   has been identified as an important
resource in many   quantum information processing  such as quantum teleportation \cite{Bennett93,Bouwmeester97} and quantum key distribution\cite{Bennett92,Gisin02}. Entanglement is a concept on   compounded quantum systems. So its existence  is  relative to the partition of the subsystems.
However,  the way to subdivide a given  physical system $S$ into
subsystems  may not be unique in general. Sometimes, the partition of subsystems depend on the physical resources available to access and manipulate the degrees of freedom of a quantum system \cite{Zana01,Zana04}.
There are some states which is entangled with respect to some subsystems but  separable with respect to another subsystems.

The concept of absolutely entangled set was proposed quite recently.  A set of bipartite states is called  an absolutely entangled set if there always exist some states in the entangled form after any action of global unitary. The study of the structure of absolutely entangled set may be helpful as  operational way
to the  device independent certification of entangled measurement \cite{Ra11}.  As a new concept, there are still many problems worthy of further study. Given a set of states, how to identify whether it is absolutely entangled or not?  How to construct small set of absolutely entangled set for general dimensional system? In this article, we will give partial answers to these questions.
 

The rest of this article is organized as follows. In Sec. \ref{second}, we give some necessary notation, and definition of the absolutely entangled set.    In Sec. \ref{third}, we study two properties on the existence of   absolutely entangled set.   In Sec. \ref{fourth}, we give two concrete constructions of absolutely entangled sets.  Finally, we draw a conclusion and present some interesting problems in     Sec. \ref{fifth}.

\section{Notation and Definition of absolutely entangled set }\label{second}
Throughout this paper, we use the following notation.
For a give positive integer $N$, denote $[N]$ to be the set $ \{1,\cdots,N\}$.    Given a nonprime integer $d$ with $d=d_1\times d_2$ where both $d_1,d_2\geq 2$ are integers, we do not distinguish the two spaces $\mathbb{C}^d$ and $\mathbb{C}^{d_1}\otimes\mathbb{C}^{d_2}$, i.e., $\mathbb{C}^d=\mathbb{C}^{d_1}\otimes\mathbb{C}^{d_2}$. And we use $\mathrm{U}(d)$ to denote the set of all unitary matrices of dimensional $d$. The states we focus here are mostly  pure states, but sometimes without normalization.

\begin{definition}  {Consider a set of pure quantum states $\{|\psi_1\rangle,...,|\psi_n\rangle\}$ in a fixed Hilbert space $\mathbb{C}^d$ of non-prime dimension. The set is said to be absolutely entangled with respect to all bipartitions into subsystems of dimension $(d_1,d_2)$, if for every unitary $U \in \mathrm{U}(d)$, at least one state $U|\psi_j\rangle$ is entangled with respect to $\mathbb{C}^{d_1} \otimes\mathbb{C}^{d_2}$.} \end{definition}

\section{Necessity condition for absolutely entangled set}\label{third}
Given a set $S$ of states in $\mathbb{C}^{d_1}\otimes \mathbb{C}^{d_2}$,  we would like to map the states in $S$ to a set of product states by a unitary matrix. The most simple set of product states is the form $|1\rangle \otimes \mathbb{C}^{d_2}\subset \mathbb{C}^{d_1}\otimes \mathbb{C}^{d_2}$. Therefore, it is interesting to ask under what condition can we embed  $S$ to the set $|1\rangle \otimes \mathbb{C}^{d_2}$. In fact, the basic linear algebra course gives us that  $S$ can be embedded to the set $|1\rangle \otimes \mathbb{C}^{d_2}$  if and only if $\mathrm{dim}_\mathbb{C} (S)\leq d_2.$ Therefore, a necessity condition for $S$ to be an absolutely entangled set is $\mathrm{dim}_\mathbb{C} (S)\geq d_2+1.$

 \begin{proposition}
 Let $d\geq 4$ be a non-prime positive integer and suppose $d=d_1d_2$ with $2\leq d_1\leq d_2$. Suppose that $S\equiv\{|\psi_j\rangle\}_{j=1}^N\subseteq \mathbb{C}^d$ is an absolutely entangled  set with respect to $(d_1,d_2)$. Then we have  
 $\emph{\text{dim}}_\mathbb{C}(S_i)\geq d_2+1$ where  $S_i\equiv S\setminus \{|\psi_i\rangle\}$ for each $i\in[N]$.

 \end{proposition}

 \noindent{ \emph{Proof}.}   Suppose not,  there exists  some $i$ such that ${\text{dim}}_\mathbb{C}(S_i)\leq d_2$.  If ${\text{dim}}_\mathbb{C}(S_i)<d_2$, then ${\text{dim}}_\mathbb{C}(S)\leq d_2$.  However, in this case, $S$  be embedded in to the subspace $|1\rangle\otimes \mathbb{C}^{d_2}$ which is contradicted with that $S$ is an absolutely entangled set.   So we can assume that ${\text{dim}}_\mathbb{C}(S_i)=d_2$.  So there exists a set of  $d_2$ orthogonal states $B\equiv\{|\xi_k\rangle\}_{k=1}^{d_2}\subseteq \mathbb{C}^d$ such that
 $$\text{span}_\mathbb{C}(S_i)=\text{span}_\mathbb{C}(B).$$
  For each $|\psi_j\rangle\in S_i$, the state $|\psi_j\rangle$ can be written as linear combination of $\{|\xi_k\rangle\}_{k=1}^{d_2}$, i.e.
 $|\psi_j\rangle=\sum_{k=1}^{d_2}c_{jk}|\xi_k\rangle.$
Any state  $|\psi\rangle \in \mathbb{C}^d$ can be written as the form $|\psi\rangle= \alpha |\xi\rangle +\beta |\eta\rangle$ where $|\xi\rangle \in  \text{span}_\mathbb{C}(S_i)$ and $|\eta\rangle \in \text{span}_\mathbb{C}(S_i)^\perp$ (the orthogonal complementary space of $\text{span}_\mathbb{C}(S_i)$).
 Specially,  we can suppose that $|\psi_i\rangle= \alpha |\xi\rangle +\beta |\eta\rangle$. As $|\xi\rangle \in \text{span}_\mathbb{C}(S_i)$,  the state $|\psi_i\rangle$ can be written as 
 $$|\psi_i\rangle=(\sum_{k=1}^{d_2}c_{ik}|\xi_k\rangle)+ \beta|\eta\rangle.$$
 
\begin{enumerate}
    \item [(i).] 
 If $c_{ik}=0$ for all $k$,  then $|\psi_i\rangle= \beta|\eta\rangle$. We can define a unitary $U\in \mathrm{U}(d)$ with the following conditions
 $U|\xi_k\rangle=|1\rangle|k\rangle, k=1,\cdots, d_2,$ and  $U|\eta\rangle= |2\rangle|1\rangle$. Obviously, under this map, the states in $S$ are being sent to product states.
 
\item [(ii).] If $c_{ik}\neq 0$ for some $k$, we can define a unitary $U$ in $\mathrm{U}(d)$ with the following conditions
 $U|\xi_k\rangle=|1\rangle|k\rangle, k=1,\cdots, d_2,$ and  $$ U|\eta\rangle=|2\rangle (\mathcal{N}[\sum_{k=1}^{d_2} c_{ik}|k\rangle])$$ where $\mathcal{N}$ is a normalized factor. Under this map, we  have
 $U|\psi_j\rangle=|1\rangle(\sum_{k=1}^{d_2}c_{jk}|k\rangle) $
 for $j\neq i$ and
 $$
 \begin{array}{ccl}
 U|\psi_i\rangle&=&|1\rangle(\displaystyle\sum_{k=1}^{d_2}c_{ik}|k\rangle) +c_{id_2+1}|2\rangle(\mathcal{N}[\sum_{k=1}^{d_2} c_{ik}|k\rangle])\\[3mm]
 &=&(|1\rangle+ \mathcal{N}  c_{id_2+1}|2\rangle)(\displaystyle\sum_{k=1}^{d_2}c_{ik}|k\rangle).
 \end{array}
 $$
 \end{enumerate} 
Both cases are contradicted with the assumption that $S$ is an absolutely entangled set.
 \qed
 
 \vskip 5pt
The above result is more general than the one proposed in Ref. \cite{Cai202006}.  In the following, we present another property on the existence of the absolutely entangled set which may not cover by the above necessity.

 \begin{proposition}
 Let $d\geq 4$ be a non-prime positive integer and suppose $d=d_1d_2$ with $2\leq d_1\leq d_2$. Let  $S\equiv\{|\psi_j\rangle\}_{j=1}^N\subseteq \mathbb{C}^d$ be a set of pure states. If $S$ can be seperated into $k$ disjoint sets $S_i\equiv \{|\psi_j^{(i)}\rangle\}_{j=1}^{N_i}$ (here $k\leq d_1$ and  $\sum_{i=1}^kN_i=N$) such that the states come from different sets are orthogonal to each other and $\emph{\text{dim}}_\mathbb{C}(S_i)\leq d_2$, then there exists some unitary $U\in\mathrm{U}(d)$ such that $\{U|\psi_j\rangle\}_{j=1}^N$ are all  in product form.
 \end{proposition}
 \noindent{ \emph{Proof}.}  Let $\{ |s\rangle|t\rangle| 1\leq s \leq d_1, 1\leq t\leq d_2\}$ denote a computational basis of the bipartite system $\mathcal{H}_A\otimes\mathcal{H}_B=\mathbb{C}^{d_1}\otimes\mathbb{C}^{d_2}$.
Denote $V_i\equiv\text{span}_\mathbb{C}(S_i)$ for $i=1,\cdots, k$ and $V\equiv\text{span}_\mathbb{C}(S)$. By the orthogonality conditions, we can conclude that there is a direct sum decomposition  $$V=\bigoplus_{i=1}^kV_i.$$
Denote $n_i\equiv\text{dim}_{\mathbb{C}}(V_i)={\text{dim}}_\mathbb{C}(S_i)$.   There exists an orthonormal basis $B_i\equiv\{|\xi_j^{(i)}\rangle\}_{j=1}^{n_i}$ of $V_i$ for each $i$. By the direct sum decomposition, we have   $B_i\cap B_j=\emptyset$ if $1\leq i\neq j\leq k$ and $\cup_{i=1}^kB_i$ is an orthonormal basis of $V$. So $\cup_{i=1}^kB_i$ can be extended to be a complete orthonormal basis of the whole space $\mathbb{C}^d$. Therefore, there exists some unitary $U\in \mathrm{U}(d)$ who maps the basis $B$ to the computational basis $\{ |s\rangle|t\rangle| 1\leq s \leq d_1, 1\leq t\leq d_2\}$. As $n_i\leq d_2$ and $k\leq d_1$, we can also assume that
$$U  |\xi_j^{(i)}\rangle=|i\rangle|j\rangle, \ \ i=1,\cdots,k, \text{ and } j=1,\cdots, n_i.$$
By the definitions of $V_i$ and $B_i$, each $|\psi_l^{(i)}\rangle\in S_i$ can be written as a linear combination of $\{|\xi_j^{(i)}\rangle\}_{j=1}^{n_i}$. That is,
$$|\psi_l^{(i)}\rangle=\sum_{j=1}^{n_i}c_{lj}^{(i)}|\xi_{j}^{(i)}\rangle, $$
$\text{ where } c_{lj}^{(i)} \in \mathbb{C}$ and $\sum_{j=1}^{n_i}|c_{lj}^{(i)}|^2=1$. Then by the definition of the unitary $U$, we have
$$U |\psi_l^{(i)}\rangle=\sum_{j=1}^{n_i}c_{lj}^{(i)}|i\rangle|j\rangle=|i\rangle(\sum_{j=1}^{n_i}c_{lj}^{(i)}|j\rangle)$$
which is a product state in the bipartite system $\mathcal{H}_A\otimes\mathcal{H}_B$.

\qed
\vskip 5pt

\section{Constructions of absolutely entangled sets}\label{fourth}

In Ref. \cite{Cai202006}, the authors presented a absolutely entangled set with four elements in $\mathbb{C}^4=\mathbb{C}^2\otimes\mathbb{C}^2$. More exactly, they proved that the set consisting of the following  four states is an absolutely entangled set
$$|\psi_1\rangle=|\xi_1\rangle, \ |\psi_i\rangle=c_{i1}|\xi_1\rangle +c_{ii}|\xi_i\rangle, i=2,3,4$$
where $\{|\xi_i\rangle\}_{i=1}^4$  is an orthonormal basis of $\mathbb{C}^4$ and $|c_{i1}|\in (\frac{1}{2},1)$ for $i=2,3,4$. Here we give an general construction of absolutely entangled set for high dimensional systems.

\begin{theorem}
Let $d\geq 4$ be a non-prime positive integer and $|\xi_1\rangle,\cdots, |\xi_d\rangle$ be an orthonormal basis of $\mathbb{C}^d$. Then  the  set consisting  of the following  $d$ states
$$\begin{array}{ll}
|\phi_1\rangle&\equiv|\xi_1\rangle;\\
|\phi_i\rangle&\equiv a_{i}|\xi_1\rangle+ b_{i}|\xi_i\rangle;\ \ \ i=2,\cdots, d
\end{array}
$$
forms an absolutely entangled set provided the conditions $\lambda|a_i|^2\geq |b_i|^2/2$ ($i=2,\cdots,d$) for some $\lambda\in (0,1)$.
\end{theorem}
\noindent{ \emph{Proof}.} Suppose not, there exists some unitary matrix $U\in\mathrm{U}(d)$ such that
$\{U|\psi_i\rangle\}_{i=1}^d$ are all product states in $\mathbb{C}^{d_1}\otimes \mathbb{C}^{d_2}$ (here we assume $d=d_1d_2$).
Without loss of generality, we assume that $U|\xi_1\rangle=|1\rangle|1\rangle$. As any unitary matrix preserves  orthogonal relations, we have
$$U|\xi_i\rangle=\sum_{(k,l)\neq (1,1)}  u_{i, c_{kl}} |k\rangle|l\rangle.$$
Here the $(u_{i, c_{kl}})$ is a $(d-1)\times(d-1)$ unitary matrix. By  definition,
$$U|\psi_i\rangle= a_i|1\rangle|1\rangle+b_i\sum_{(k,l)\neq (1,1)} u_{i, c_{kl}} |k\rangle|l\rangle, \ \text{ for } i=2,\cdots,d$$
are all in product form.
Therefore, each of the following matrix
$$
\left[\begin{array}{ccccc}
a_{i}& b_i u_{i, c_{12}}&b_i u_{i, c_{13}}&\cdots& b_i u_{i, c_{1d_2}}\\
b_i u_{i, c_{21}}& b_i u_{i, c_{22}}& b_i u_{i, c_{23}}&\cdots& b_i u_{i, c_{2d_2}}\\
b_i u_{i, c_{31}}& b_i u_{i, c_{32}}& b_i u_{i, c_{33}}&\cdots& b_i u_{i, c_{3d_2}}\\
\cdots& \vdots& \vdots&\cdots& \vdots \\
b_i u_{i, c_{d_11}}& b_i u_{i, c_{d_12}}& b_i u_{i, c_{d_13}}&\cdots& b_i u_{i, c_{d_1d_2}}\\
\end{array}\right]
$$
is   rank  one. Particularly,  the left top $2\times 2 $ minor  must be zero. That is,
$a_ib_i u_{i, c_{22}}-b_i u_{i, c_{12}}b_i u_{i, c_{21}}=0$  which is equivalent to
$a_i u_{i, c_{22}}= b_i u_{i, c_{12}} u_{i, c_{21}}$ as $b_i\neq 0$.  By Cauchy inequality,
$$|a_i|\cdot | u_{i, c_{22}}|=|b_i| \cdot|u_{i, c_{12}}| \cdot|u_{i, c_{21}}|\leq |b_i| \frac{|u_{i, c_{12}}|^2+|u_{i, c_{21}}|^2}{2}.$$
Squaring both sides of the above inequality, we obtain
$$\begin{array}{rcl}
|a_i|^2\cdot | u_{i, c_{22}}|^2&\leq& |b_i|^2 (\frac{|u_{i, c_{12}}|^2+|u_{i, c_{21}}|^2}{2})^2\\[2mm]
                       &\leq&\frac{|b_i|^2}{4} (|u_{i, c_{12}}|^2+|u_{i, c_{21}}|^2).
\end{array}
$$
The second inequality holds because of $|u_{i, c_{12}}|^2+|u_{i, c_{21}}|^2\leq 1$. As $\lambda|a_i|^2>|b_i|^2/2$ holds by assumption,   the above inequality implies that
$$| u_{i, c_{22}}|^2\leq\lambda(|u_{i, c_{12}}|^2+|u_{i, c_{21}}|^2)/2$$
for $i=2,\cdots, d$. Summing up the two sides of $i$ from 2 to $d$,
we obtain
$$1=\sum_{i=2}^d| u_{i, c_{22}}|^2\leq\lambda(\sum_{i=2}^d|u_{i, c_{12}}|^2+\sum_{i=2}^d|u_{i, c_{21}}|^2)/2=\lambda<1.$$
Wherehence, we obtain a contradiction. So the original set must be absolutely entangled.  \qed

\vskip 5pt

In the following, we will present another example of absolutely entangled set in $\mathbb{C}^4$. However, the number of elements of our set here is five instead of four.

\begin{example}\label{examone}
 Let $d=4$, $p=2$, and  $\{|\xi_i\rangle\}_{i=1}^4$ be an orthonormal basis of $\mathbb{C}^4$. Then the   set consisting of the following five states
 {\small$$\begin{array}{ll}
 |\psi_i\rangle=|\xi_i\rangle, & i=1,2,3,\\[2mm]
 |\psi_4\rangle=\displaystyle\sum_{j=1}^4x^{p^j}|\xi_j\rangle, &
 |\psi_{5}\rangle=\displaystyle\sum_{j=1}^4x^{2p^j}|\xi_j\rangle
 \end{array}
 $$
 }
 is an absolutely entangled set for almost all $x\in (0,1)$ (That is, there are  only   finitely exceptional cases).
 \end{example}

\noindent{\emph{Proof.}} Suppose not,  there is a unitary transformation $U\in\mathrm{U}(4)$ such that $\{U|\psi_i\rangle\}_{i=1}^{5}$ are all in product form of $\mathcal{H}_A\otimes\mathcal{H}_B=\mathbb{C}^2\otimes\mathbb{C}^2$. Particularly,
$\{U|\xi_i\rangle\}_{i=1}^{3}$ are orthonormal product states in $\mathcal{H}_A\otimes\mathcal{H}_B$.  As any unitary preserves the orthogonality,  $U|\xi_4\rangle$ should be also orthogonal to all $\{U|\xi_i\rangle\}_{i=1}^{3}$. As it is well known that there is no UPB whose complementary space is of dimensional one.  That is, any set of orthogonal product states with $d-1$ elements can be extended to an orthogonal product basis. However, there is only one  vector  which is orthogonal  $\{U|\xi_i\rangle\}_{i=1}^{3}$ up to a global phase.  Hence  $U|\xi_4\rangle$ must also be in a product form. So the unitary $U$ maps the basis $\{|\xi_i\rangle\}_{i=1}^4$ to an orthonormal product basis of  $\mathcal{H}_A\otimes\mathcal{H}_B$.
It is not difficult to show that the general orthonormal product basis of  $\mathcal{H}_A\otimes\mathcal{H}_B$ is of the form (or an exchange the two particles)
$$|\alpha\rangle|\beta\rangle, |\alpha^\perp\rangle|\gamma\rangle,|\alpha\rangle|\beta^\perp\rangle,|\alpha^\perp\rangle|\gamma^\perp\rangle.$$

Then $U|\psi_4\rangle $ and $U|\psi_{5}\rangle$ can be written as
$$
\begin{array}{l}
|\alpha\rangle (x^{p^i}|\beta\rangle +x^{p^j}|\beta^\perp\rangle) +|\alpha^\perp\rangle (x^{p^k}|\gamma\rangle +x^{p^l}|\gamma^\perp\rangle), \\[2mm]
|\alpha\rangle (x^{2p^i}|\beta\rangle +x^{2p^j}|\beta^\perp\rangle) +|\alpha^\perp\rangle (x^{2p^k}|\gamma\rangle +x^{2p^l}|\gamma^\perp\rangle) .
\end{array}
$$
respectively. One necessary condition for the above two states to be in product form is that
\begin{equation}\label{para}
\begin{array}{l}
x^{p^i}|\beta\rangle +x^{p^j}|\beta^\perp\rangle// x^{p^k}|\gamma\rangle +x^{p^l}|\gamma^\perp\rangle;\\
x^{2p^i}|\beta\rangle +x^{2p^j}|\beta^\perp\rangle// x^{2p^k}|\gamma\rangle +x^{2p^l}|\gamma^\perp\rangle.
\end{array}
\end{equation}
Here $i,j,k,l$ is a complete list  of the four numbers $1,2,3,4$. Suppose that
$$\begin{array}{l}
|\gamma\rangle=v_{11}|\beta\rangle+v_{12}|\beta^\perp\rangle, \ \ \ 
|\gamma^\perp\rangle=v_{12}|\beta\rangle+v_{22}|\beta^\perp\rangle.
\end{array}
$$
 Then the matrix $V\equiv (v_{ij})$ is a unitary matrix. The relations in Eq. (\ref{para}) are equivalent to  that
$$\mathcal{N}([x^{p^{k}},x^{p^{l}}])V=e^{i\theta_1}\mathcal{N}([x^{p^{i}},x^{p^{j}}])$$
and
$$\mathcal{N}([x^{2p^{k}},x^{2p^{l}}])V=e^{i\theta_2}\mathcal{N}([x^{2p^{i}},x^{2p^{j}}]).$$
Here  and below we use the notation $\mathcal{N}(v)$ to denote the vector  $\frac{v}{\|v\|}$ for any nonzero vector $v$.

Noticing that any unitary matrix should preserve the inner product of vectors. As $x$ is a real number by assumption, the following two terms
{\small $$\frac{ x^{3p^{k}}+x^{3p^{l}}}{\sqrt{x^{2p^{k}}+x^{2p^{l}}}\sqrt{x^{4p^{k}}+x^{4p^{l}}}}, \ \ \text{ and }\
 \frac{ x^{3p^{i}}+x^{3p^{i}}}{\sqrt{x^{2p^{i}}+x^{2p^{j}}}\sqrt{x^{4p^{i}}+x^{4p^{j}}}}$$}
are equal. This is equivalent to the  equation
{\small\begin{equation}
\frac{ (x^{3p^{k}}+x^{3p^{l}})^2}{({x^{2p^{k}}+x^{2p^{l}}})({x^{4p^{k}}+x^{4p^{l}}})}=
 \frac{ (x^{3p^{i}}+x^{3p^{j}})^2}{({x^{2p^{i}}+x^{2p^{j}}})({x^{4p^{i}}+x^{4p^{j}}})}.
 \end{equation}
 }
We define $f_{ij|kl}^{(p)}(X)$ as the following
{\small $$\begin{array}{rcl} f_{ij|kl}^{(p)}(X)&\equiv&(X^{3p^{k}}+X^{3p^{l}})^2({X^{2p^{i}}+X^{2p^{j}}})({X^{4p^{i}}+X^{4p^{j}}})\\[2mm]
& &-(X^{3p^{i}}+X^{3p^{j}})^2({X^{2p^{k}}+X^{2p^{l}}})({X^{4p^{k}}+X^{4p^{l}}}).
\end{array}
$$
}
 Using mathematica, we can calculate the three expressions $f_{12|34}^{(2)}(X)$, $f_{13|24}^{(2)}(X),$ and  $f_{14|23}^{(2)}(X)$ respectively as follows:
\begin{widetext}
{\small$$\begin{array}{l}
 -X^{64} + 2 X^{66} - X^{68} + X^{76} + 2 X^{82} - 2 X^{84} - X^{88} - X^{92} -
 2 X^{96} + 2 X^{98} + X^{104} - X^{112} + 2 X^{114} - X^{116},\\[2mm]
 -X^{48} + 2 X^{54} - 2 X^{72} + 2 X^{78} - X^{84} - X^{96} + 2 X^{102} - 2 X^{108} +
 2 X^{126} - X^{132},\\[2mm]
X^{44} - 2 X^{48} + X^{52} - X^{64} - 2 X^{76} + 2 X^{78} +
 2 X^{86} - X^{88} - X^{92} + 2 X^{94} + 2 X^{102} - 2 X^{104} - X^{116} + X^{128} -
 2 X^{132} + X^{136}.
 \end{array}
 $$
 }
\begin{table}[h]
\caption{ The real roots of the above three polynomials for constructing absolutely entangled set in $\mathbb{C}^4$.}\label{dimensionalfour}
\begin{center}
	\begin{tabular}{c|ccccccccccccc}\toprule[2pt]
			\diagbox{{Poly}}{Roots}{Num} & 1&&2&&3 &&4&&5 &&6&	&7\\\hline
	   $f_{12|34}^{(2)}(X)$& -1.21341 & &-1&& -0.824127 & &0  && 0.824127  && 1&&1.21341  \\[2mm]
			$f_{13|24}^{(2)}(X)$ & -1.10104  && -1 && -0.908231 && 0 && 0.908231   & &1&&1.10104 \\[2mm]
				$f_{14|23}^{(2)}(X)$ & -1.11046   & &-1  && -0.900525   && 0 && 0.900525 & &1&&1.11046 \\ \bottomrule[2pt]
			\end{tabular}

\end{center}
\end{table}
 \end{widetext}
   There are only fifteen real numbers $x_i$ (see Table \ref{dimensionalfour}) s. t.
   $$f_{12|34}^{(2)}(x_i)f_{13|24}^{(2)}(x_i)f_{14|23}^{(2)}(x_i)=0.$$
 So for any real number $x\neq x_i$ for all $i=1,\cdots,15$, the set consisting of  the five states we constructed above   indeed forms an absolutely entangled set. \qed

\vskip 5pt

In fact, we idea of Example \ref{examone} can be extended to the bipartite system $\mathbb{C}^2\otimes\mathbb{C}^n$. But we need some more results on the   structure of  product basis in the system of $\mathbb{C}^2\otimes\mathbb{C}^n$.   Based on the  structure results,   we will  present another construction of absolutely entangled set of the form $(2,n)$. Now we present the statement of some known results about orthnormal product basis in $\mathbb{C}^2\otimes\mathbb{C}^n$.

\vskip 5pt

\begin{lemma}[Feng 2006, see Ref. \cite{Feng06}]\label{Feng}
Let  $\mathcal{H}_A=\mathbb{C}^2$ and $\mathcal{H}_B=\mathbb{C}^n$. Any orthonormal product basis of $\mathcal{H}_A\otimes\mathcal{H}_B$  can be written in the following way:
 $$
 S=\{|a_k\rangle\otimes \mathcal{A}_k, |a_k^\perp\rangle\otimes \mathcal{A}_k' \ \big | \ 1\leq k \leq t\}$$
where $\{|a_k\rangle, |a_k^\perp\rangle\}$   are different (normalized) orthogonal basis in $\mathcal{H}_A$ and $\mathcal{A}_k=\{|u^{(k)}_i\rangle\}_{i=1}^{n_k}, \mathcal{A}'_k=\{|v^{(k)}_i\rangle\}_{i=1}^{n'_k}\subseteq  \mathcal{H}_B$ ($1\leq k \leq t$) such that the following conditions hold
\begin{enumerate}[(1)]
  \item  $\langle u^{(k)}_i|u^{(k)}_j\rangle= \delta_{ij}$;
  \item  $\langle v^{(k)}_i|v^{(k)}_j\rangle= \delta_{ij}$;
  \item  $n_k=\text{\emph{span}}_\mathbb{C}(\mathcal{A}_k)=\text{\emph{span}}_\mathbb{C}(\mathcal{A}_k')= n_k';$
  \item $\mathcal{H}_B=\bigoplus_{k=1}^t \text{\emph{span}}_\mathbb{C}(\mathcal{A}_k).$
\end{enumerate}

\end{lemma}

\begin{theorem}
 Let $d\geq 4$ be an even integer and suppose $d=2n$. Let $\{|\xi_i\rangle\}_{i=1}^d$ be an orthonormal basis of $\mathbb{C}^d$. Then the  set consisting of the following $d+1$ states
 {\small$$\begin{array}{ll}
 |\psi_i\rangle=|\xi_i\rangle, & i=1,2,\cdots, d-1,\\[2mm]
 |\psi_d\rangle=\displaystyle\sum_{j=1}^dx^{p^j}|\xi_j\rangle,&
 |\psi_{d+1}\rangle=\displaystyle\sum_{j=1}^dx^{2p^j}|\xi_j\rangle
 \end{array}
 $$
 }
 is an absolutely entangled set of the form $(2,n)$ for almost all $x\in (0,1)$ provided $p$ to be an integer greater than 7.
 \end{theorem}

\noindent{\emph{Proof.}} Suppose not,  there is a unitary transformation $U\in\mathrm{U}(d)$ such that $\{U|\psi_i\rangle\}_{i=1}^{d+1}$ are all in product form of $\mathcal{H}_A\otimes\mathcal{H}_B=\mathbb{C}^2\otimes\mathbb{C}^n$. Particularly,
$\{U|\xi_i\rangle\}_{i=1}^{d-1}$ are orthonormal product states in $\mathcal{H}_A\otimes\mathcal{H}_B$.  As any unitary preserves the orthogonality,  $U|\xi_d\rangle$ should be orthogonal to all $\{U|\xi_i\rangle\}_{i=1}^{d-1}$. Similar with the argument of  Example \ref{examone},  the unitary $U$ maps the basis $\{|\xi_i\rangle\}_{i=1}^d$ to an orthonormal product basis of  $\mathcal{H}_A\otimes\mathcal{H}_B$. By Lemma \ref{Feng}, any orthonormal product basis of $\mathcal{H}_A\otimes\mathcal{H}_B$ is of the form
$$
 S=\{|a_k\rangle\otimes \mathcal{A}_k, |a_k^\perp\rangle\otimes \mathcal{A}_k' \ \big | \ 1\leq k \leq t\}$$
 where $|a_i\rangle,|a_i^\perp\rangle,\mathcal{A}_i,\mathcal{A}_i'$ satisfy  the four conditions in the statement of Lemma \ref{Feng}.

Then $U|\psi_d\rangle $ and $U|\psi_{d+1}\rangle$ can be written as
{\small\begin{equation}\label{two_orthogonal_states}
\begin{array}{l}
U|\psi_d\rangle=\displaystyle\sum_{k=1}^t(\sum_{i=1}^{n_k}x^{p^{g_{ki}}}|a_k\rangle|u_i^{(k)}\rangle+
\sum_{i=1}^{n_k}x^{p^{h_{ki}}}|a_k^\perp\rangle|v_i^{(k)}\rangle),\\[2mm]
U|\psi_{d+1}\rangle=\displaystyle\sum_{k=1}^t(\sum_{i=1}^{n_k}x^{2p^{g_{ki}}}|a_k\rangle|u_i^{(k)}\rangle+
\sum_{i=1}^{n_k}x^{2p^{h_{ki}}}|a_k^\perp\rangle|v_i^{(k)}\rangle).
\end{array}
\end{equation}
}
Here $g_{ki},h_{ki}, (k=1,\cdots,t, i=1,\cdots,n_k)$ is a complete list of $1,2,3,\cdots,d$.  We separate the analysis into two cases: (1) all $n_k$ are equal to 1; (2) at least one $n_k\geq 2$.

  (1) In this setting, $t=n$, $|u_1^{(k)}\rangle=|v_1^{(k)}\rangle$  up to a phase and $\{|u_1^{(k)}\rangle\}_{k=1}^n$ form an orthonormal basis of $\mathcal{H}_B$. Eqs. (\ref{two_orthogonal_states}) can be simplified as

  \begin{equation}\label{simple_two_orthogonal_states}
\begin{array}{l}
U|\psi_d\rangle=\displaystyle\sum_{k=1}^n ( x^{p^{g_{k1}}}|a_k\rangle+
 x^{p^{h_{k1}}}|a_k^\perp\rangle)|u_1^{(k)}\rangle,\\[2mm]
U|\psi_{d+1}\rangle=\displaystyle\sum_{k=1}^n(x^{2p^{g_{k1}}}|a_k\rangle +
 x^{2p^{h_{k1}}}|a_k^\perp\rangle)|u_1^{(k)}\rangle.
\end{array}
\end{equation}
  One necessary condition for the two states in Eqs. (\ref{simple_two_orthogonal_states})  to be in product form is that for $k\neq t$
$$
\begin{array}{c}
 x^{p^{g_{k1}}}|a_k\rangle+
 x^{p^{h_{k1}}}|a_k^\perp\rangle//  x^{p^{g_{t1}}}|a_t\rangle+
 x^{p^{h_{t1}}}|a_t^\perp\rangle;\\
 x^{p^{g_{2k1}}}|a_k\rangle+
 x^{p^{2h_{k1}}}|a_k^\perp\rangle//  x^{p^{2g_{t1}}}|a_t\rangle+
 x^{p^{2h_{t1}}}|a_t^\perp\rangle.
\end{array}
$$
As $g_{k1},h_{k1},g_{t1},h_{t1}$ are four different elements in $\{1,2,\cdots,d\}$, we use $i,j,k,l$ to denote this four elements respectively for simplicity. Under this simplification, the above relations are  equivalent to  that there exists an $2\times 2$ unitary matrix $V$ of  such that
$$\mathcal{N}([x^{p^{k}},x^{p^{l}}])V=e^{i\theta_1}\mathcal{N}([x^{p^{i}},x^{p^{j}}])$$
and
$$\mathcal{N}([x^{2p^{k}},x^{2p^{l}}])V=e^{i\theta_2}\mathcal{N}([x^{2p^{i}},x^{2p^{j}}]).$$
And the unitary matrix should preserve the inner product of vectors. As $x$ is a real number by assumption, the following two terms
{\small $$\frac{ x^{3p^{k}}+x^{3p^{l}}}{\sqrt{x^{2p^{k}}+x^{2p^{l}}}\sqrt{x^{4p^{k}}+x^{4p^{l}}}}, \ \ \text{ and }\
 \frac{ x^{3p^{i}}+x^{3p^{i}}}{\sqrt{x^{2p^{i}}+x^{2p^{j}}}\sqrt{x^{4p^{i}}+x^{4p^{j}}}}$$}
are equal. This is equivalent to the  equation
\begin{equation}\label{function2n_2}
\frac{ (x^{3p^{k}}+x^{3p^{l}})^2}{({x^{2p^{k}}+x^{2p^{l}}})({x^{4p^{k}}+x^{4p^{l}}})}=
 \frac{ (x^{3p^{i}}+x^{3p^{j}})^2}{({x^{2p^{i}}+x^{2p^{j}}})({x^{4p^{i}}+x^{4p^{j}}})}.
 \end{equation}
We define $f_{ij|kl}^{(p)}(X)$ as the following
{\small $$\begin{array}{rcl} f_{ij|kl}^{(p)}(X)&\equiv&(X^{3p^{k}}+X^{3p^{l}})^2({X^{2p^{i}}+X^{2p^{j}}})({X^{4p^{i}}+X^{4p^{j}}})\\[2mm]
& &-(X^{3p^{i}}+X^{3p^{j}})^2({X^{2p^{k}}+X^{2p^{l}}})({X^{4p^{k}}+X^{4p^{l}}}).
\end{array}
$$
}
Therefore, in this case, the $x$ should be a root of  $f_{ij|kl}^{(p)}(X)=0$ by Eq. (\ref{function2n_2}).

(2) Suppose that $n_k\geq 2$ for some $k$. One necessary condition for the two states in Eqs. (\ref{two_orthogonal_states})  to be in product form is that

$$
\begin{array}{c}
\displaystyle\sum_{i=1}^{n_k}x^{p^{g_{ki}}}|u_i^{(k)}\rangle// \sum_{i=1}^{n_k}x^{p^{h_{ki}}}|v_i^{(k)}\rangle;\\
\displaystyle\sum_{i=1}^{n_k}x^{2p^{g_{ki}}}|u_i^{(k)}\rangle// \sum_{i=1}^{n_k}x^{2p^{h_{ki}}}|v_i^{(k)}\rangle.
\end{array}
$$
This is equivalent to  that there exists an $n_k\times n_k$ unitary matrix $W$   such that
$$\mathcal{N}([x^{p^{h_{k1}}},\cdots,x^{p^{h_{kn_k}}}])W=e^{i\phi_1}\mathcal{N}([x^{p^{g_{k1}}},\cdots,x^{p^{g_{kn_k}}}])$$
and
$$\mathcal{N}([x^{2p^{h_{k1}}},\cdots,x^{2p^{h_{kn_k}}}])W=e^{i\phi_2}\mathcal{N}([x^{2p^{g_{k1}}},\cdots,x^{2p^{g_{kn_k}}}]).$$
And the unitary matrix $W$ should preserve the inner product of the two vectors. That is,
{\small$$\frac{\displaystyle\sum_{i=1}^{n_k} x^{3p^{h_{ki}}}}{\sqrt{\displaystyle\sum_{i=1}^{n_k} x^{2p^{h_{ki}}}}\sqrt{\displaystyle\sum_{i=1}^{n_k} x^{4p^{h_{ki}}}}}, \ \ \text{ and }\
 \frac{\displaystyle\sum_{i=1}^{n_k} x^{3p^{g_{ki}}}}{\sqrt{\displaystyle\sum_{i=1}^{n_k} x^{3p^{g_{ki}}}}\sqrt{\displaystyle\sum_{i=1}^{n_k} x^{4p^{g_{ki}}}}}$$
 }
are equal. This is equivalent to the following equation
{\small\begin{equation}\label{function2n_general}
\frac{(\displaystyle\sum_{i=1}^{n_k} x^{3p^{h_{ki}}})^2}{{\displaystyle\sum_{i=1}^{n_k} x^{2p^{h_{ki}}}}{\displaystyle\sum_{i=1}^{n_k} x^{4p^{h_{ki}}}}}=\frac{(\displaystyle\sum_{i=1}^{n_k} x^{3p^{g_{ki}}})^2}{{\displaystyle\sum_{i=1}^{n_k} x^{2p^{g_{ki}}}}{\displaystyle\sum_{i=1}^{n_k} x^{4p^{g_{ki}}}}}.
\end{equation}
}
We define $f_{h_{k1}\ldots h_{k n_k}|g_{k1}\ldots g_{k n_k}}^{(p)}(X)$ as the following

{\small $$
\begin{array}{rcl}
f_{h_{k1}\ldots h_{k n_k}|g_{k1}\ldots g_{k n_k}}^{(p)}(X)&=&(\displaystyle\sum_{i=1}^{n_k} X^{3p^{h_{ki}}})^2{\displaystyle\sum_{i=1}^{n_k} X^{2p^{g_{ki}}}}{\displaystyle\sum_{i=1}^{n_k} X^{4p^{g_{ki}}}}\\
&&-(\displaystyle\sum_{i=1}^{n_k} X^{3p^{g_{ki}}})^2{\displaystyle\sum_{i=1}^{n_k} X^{2p^{h_{ki}}}}{\displaystyle\sum_{i=1}^{n_k} X^{4p^{h_{ki}}}}.
\end{array}$$
}
By Eq. (\ref{function2n_general}), $x$ should be a root of the equation $$f_{h_{k1}\ldots h_{k n_k}|g_{k1}\ldots g_{k n_k}}^{(p)}(X)=0.$$

For each integer $s$ ($2\leq s\leq n$), define $\mathcal{P}_s$ to be the set
{\small$$\{ f^{(p)}_{i_1\ldots i_s|j_1\ldots j_s}(X) \big |  i_1,\ldots, i_s,j_1,\ldots, j_s \in [2n]  \text{ and are different }\}$$ }
\noindent where $f^{(p)}_{i_1\ldots i_s|j_1\ldots j_s}(X)$ is defined as the Lemma \ref{nonzerolemma}. And we denote $\mathcal{P}:=\bigcup_{s=2}^n \mathcal{P}_s.$

 To make all the  states in the set we construct to be separable, the $x$ must be a root of some polynomial in $\mathcal{P}$. Now set $$ \mathcal{X}:=\{x\in (0,1) | \exists f \in \mathcal{P}, \text{ s.t. }  f(x)=0 \}.$$  However, there are only a finite number  of   polynomials  in $\mathcal{P}$ and each is nonzero by Lemma \ref{nonzerolemma}.  Therefore, $\mathcal{X}$ is a finite set and any $x\in (0,1)\setminus \mathcal {X}$ do not satisfy the Eqs. (\ref{function2n_2}) and (\ref{function2n_general}).  For such $x$, the set of $d+1$ states we constructed  is an absolutely entangled set.
 \qed

\begin{lemma}\label{nonzerolemma}
Let $s\geq 2$ and $p\geq 7$ be  integers. Given any $2s$ different positive integers $h_1,\cdots,h_s, g_1,\cdots,g_s$,
we define a polynomial $f_{h_{1}\ldots h_{s}|g_{1}\ldots g_{s}}^{(p)}(X)$ as the following

{\small$$(\displaystyle\sum_{i=1}^{s} X^{3p^{h_{i}}})^2{\displaystyle\sum_{k=1}^{s} X^{2p^{g_{k}}}}{\displaystyle\sum_{l=1}^{s} X^{4p^{g_{l}}}}-(\displaystyle\sum_{k=1}^{s} X^{3p^{g_{k}}})^2{\displaystyle\sum_{i=1}^{s} X^{2p^{h_{i}}}}{\displaystyle\sum_{j=1}^{s} X^{4p^{h_{j}}}}.$$
}
Then $f_{h_{1}\ldots h_{s}|g_{1}\ldots g_{s}}^{(p)}(X)$ is a nonzero polynomial.
\end{lemma}
\noindent \emph{Sketch of proof.} The first term and second term of $f_{h_{1}\ldots h_{s}|g_{1}\ldots g_{s}}^{(p)}(X)$ can be expanded as
\begin{equation}\label{two_terms}
\begin{array}{l}
\displaystyle\sum_{i=1}^s\sum_{j=1}^s\sum_{k=1}^s\sum_{l=1}^s X^{3p^{h_i}+3p^{h_j}+2p^{g_k}+4p^{g_l}},\\
\displaystyle\sum_{i'=1}^s\sum_{j'=1}^s\sum_{k'=1}^s\sum_{l'=1}^s X^{2p^{h_{i'}}+4p^{h_{j'}}+3p^{g_{k'}}+3p^{g_{l'}}}
\end{array}
\end{equation}
respectively. As $p\geq 7$ and the difference of $h_1,\cdots,h_s, g_1,\cdots,g_s$, the following equality holds
$$3p^{h_i}+3p^{h_j}+2p^{g_k}+4p^{g_l}=2p^{h_{i'}}+4p^{h_{j'}}+3p^{g_{k'}}+3p^{g_{l'}}$$
if and only if $i=j=i'=j'$ and $k=l=k'=l'$. Therefore, only $s^2$ terms in the first expansion   of Eq. (\ref{two_terms}) can be eliminated from the terms in the second expansion.   From this, we can conclude that $f_{h_{1}\ldots h_{s}|g_{1}\ldots g_{s}}^{(p)}(X)$ is  indeed a nonzero polynomial. \qed

\vskip 5pt

\section{Conclusion and Discussion}\label{fifth}
We  have derived two necessary conditions for a set to be an absolutely entangled set.  This gives an improvement of that obtained in the original paper of  absolutely entangled set \cite{Cai202006}. We also presented  two constructions of  the absolutely entangled sets on the systems
 $\mathbb{C}^m\otimes\mathbb{C}^n$   and    $\mathbb{C}^2\otimes\mathbb{C}^n$    respectively.  The first construction is  inspired by the construction of    Cai et al.'s results and  it holds for general dimensional systems. The second one is more technical but only holds in special dimensional systems.
 
 As there is no orthogonal product basis structure for general bipartite systems, we do not know whether our second construction can be extended to more general dimensional cases.   And we do not consider the  optimal  construction in this paper. However, it is much more interesting to find out the smallest sets of absolutely entangled set of a given dimension.
 
 \vskip 5pt
 

\noindent{\bf Acknowledgments}\, \,  This  work  is supported  by  National  Natural  Science  Foundation  of  China  (11875160, 12005092  and  U1801661),  the China Postdoctoral Science Foundation (2020M681996), the  Natural  Science  Foundation  of  Guang-dong  Province  (2017B030308003),  the  Key  R$\&$D  Program of   Guangdong   province   (2018B030326001),   the   Guang-dong    Innovative    and    Entrepreneurial    Research    TeamProgram (2016ZT06D348), the Science, Technology and   Innovation   Commission   of   Shenzhen   Municipality (JCYJ20170412152620376   and   JCYJ20170817105046702 and  KYTDPT20181011104202253),    the Economy, Trade  and  Information  Commission  of  Shenzhen Municipality (201901161512).

\end{document}